\documentclass[lettersize,journal]{IEEEtran}
\usepackage{amsmath,amsfonts}
\usepackage{algorithmic}
\usepackage{subfigure}
\usepackage{setspace}
\usepackage{algorithm}
\usepackage{array}
\usepackage{textcomp}
\usepackage{stfloats}
\usepackage{url}
\usepackage{verbatim}
\usepackage{booktabs}
\usepackage{graphicx}
\usepackage[table,x11names]{xcolor}
\usepackage{cite}
\usepackage{caption}
\usepackage{subcaption}
\usepackage{svg}
\usepackage{hyperref}
\usepackage{xspace}
\usepackage{scalerel}
\usepackage{tikz}
\usetikzlibrary{svg.path}

\definecolor{orcidlogocol}{HTML}{A6CE39}
\tikzset{
  orcidlogo/.pic={
    \fill[orcidlogocol] svg{M256,128c0,70.7-57.3,128-128,128C57.3,256,0,198.7,0,128C0,57.3,57.3,0,128,0C198.7,0,256,57.3,256,128z};
    \fill[white] svg{M86.3,186.2H70.9V79.1h15.4v48.4V186.2z}
                 svg{M108.9,79.1h41.6c39.6,0,57,28.3,57,53.6c0,27.5-21.5,53.6-56.8,53.6h-41.8V79.1z M124.3,172.4h24.5c34.9,0,42.9-26.5,42.9-39.7c0-21.5-13.7-39.7-43.7-39.7h-23.7V172.4z}
                 svg{M88.7,56.8c0,5.5-4.5,10.1-10.1,10.1c-5.6,0-10.1-4.6-10.1-10.1c0-5.6,4.5-10.1,10.1-10.1C84.2,46.7,88.7,51.3,88.7,56.8z};
  }
}

\newcommand\orcidicon[1]{\href{https://orcid.org/#1}{\mbox{\scalerel*{
\begin{tikzpicture}[yscale=-1,transform shape]
\pic{orcidlogo};
\end{tikzpicture}
}{|}}}}

\begin{document}

\title{Configurable Multi-Port Memory Architecture for High-Speed Data Communication}

\author{Narendra Singh Dhakad \orcidicon{0000-0003-2848-1785} and Santosh Kumar Vishvakarma \orcidicon{0000-0003-4223-0077}
\\ \vspace{0.5em} \small{Indian Institute of Technology Indore, India}
}

\maketitle
\begin{abstract}
Memory management is necessary with the increasing number of multi-connected AI devices and data bandwidth issues. For this purpose, high-speed multi-port memory is used. The traditional multi-port memory solutions are hard-bounded to a fixed number of ports for read or write operations. In this work, we proposed a pseudo-quad-port memory architecture. Here, ports can be configured (1-port, 2-port, 3-port, 4-port) for all possible combinations of read/write operations for the 6T static random access memory (SRAM) memory array, which improves the speed and reduces the bandwidth for data transfer. The proposed architecture improves the bandwidth of data transfer by 4$\times$. The architecture provides 1.3$\times$ and 2$\times$ area efficiency compared to dual-port 8T and quad-port 12T SRAM. 
\end{abstract}

\begin{IEEEkeywords}
Multi-port memory, memory addressing, high-bandwidth memory
\end{IEEEkeywords}

\section{Introduction}
\IEEEPARstart {M}emory plays a crucial role in electronic systems. With the advancement of edge AI devices, efficient memory management is essential. Due to their complex computational needs, graphic, audio, and video processing require heterogeneous and multi-core processors. Multi-port memories have become vital for CPUs since multi-core processors demand significant data transfer capabilities. For edge AI applications such as smartwatches and autonomous cars, which receive or transfer data from various sources, reconfigurable multi-port memories are necessary for fast and efficient design. Distributed arithmetic-based architectures also require multi-port SRAM circuits for high throughput design \cite{R1}.

However, multi-port memories come with increased area and leakage costs. Traditionally, dual-port SRAMs use an 8T SRAM bitcell, which has a high area and leakage and faces contention issues when read word line (RWL) and write word line (WWL) is selected for the same bitcell. The write driver can disturb stored data by driving the write bitline. Similar problems can occur in most conventional dual-port memories, such as 8T, 10T, and 12T configurations. Additionally, this memory architecture can only function as a two-port memory, with one port fixed for reading and the other for writing. \cite{dualportsurvey, dhakad, 1R1W}.

This paper presents a wrapper circuit that transforms a conventional single-port 6T SRAM macro into a multi-port configuration. The architecture offers flexibility, allowing it to be configured for 1-port, 2-port, 3-port, or 4-port operations. The proposed approach avoids contention issues by accessing the cell sequentially, enhancing read and write robustness. Additionally, the ports can be configured for various combinations of read and write operations, such as 1-read/3-write, 2-read/2-write, 3-read/1-write, or all read/write ports. Utilizing 6T SRAM bit cells increases the memory array density, improving area efficiency. Integrating a wrapper for latches and multiplexers around a regular single-port SRAM enables pseudo-multi-port operation.

\begin{table*}[t]
\caption{\scshape Available multi-port Architectures}
\centering
\begin{tabular}{|l|c|c|c|c|c|c|c|c|} \hline
\rowcolor{lightgray}\textbf{Parameters} & \textbf{1R1W} \cite{1R1W} & \textbf{2R2W} \cite{2R2W} & \textbf{8R1W} \cite{1W/8R} & \textbf{5R1W} \cite{5r1w} & \textbf{6R2W} \cite{6r2w}  & \textbf{6R6W} \cite{6r6w} & \textbf{4R1W} \cite{AlgorithmicMultiPort}    &  \textbf{Multi-port} \cite{MultiPortCoding} \\ \hline
\cellcolor{lightgray}Bitcell  &  8T  &  12T  &  20T  &  16T  &  24T  &  16T   & -- & DRAM \\ \hline
\cellcolor{lightgray}\#N Ports   & 2   & 4   & 9   & 6    & 8   & 12    & 5 & N \\ \hline
\cellcolor{lightgray}Ports  & Fixed  & Fixed & Fixed  & Fixed &  Fixed & Fixed  & Configurable & Configurable\\ \hline
\cellcolor{lightgray}Approach & Bitcell & Bitcell  & Bitcell  & Bitcell  & Bitcell  &  Bitcell & Algorithm & Coding \\\hline
\end{tabular}
\label{literature}
\end{table*}

A simpler way to add multiple ports is to modify the memory cell with additional wordlines and bitlines. Many works \cite{dualportsurvey, dhakad, 1R1W} have been proposed for dual port work, where one port is used for read and another for write operation and \cite{1W/8R, 6r2w, 6r6w,5r1w,2R2W} presented multi-port memory designs. However, these ports are hard-bounded and can not change operations once fabricated. Also, such dual-port SRAMs use a higher transistor SRAM bitcell, which increases the area and leakage and also suffers from contention issues.

An algorithmic multi-port memory configuration is presented in \cite{AlgorithmicMultiPort}, which schedules multi-port memory instructions in the framework by extracting parallelism from the algorithm trace. Similarly, a coding-based solution is proposed in \cite{MultiPortCoding}. These approaches are effective for field programmable gate arrays (FPGAs) but cannot be deployed for memory chip design.

2R2W \cite{2R2W} proposed a four-port 12T memory cell that can perform two red and two write operations. However, it has an area overhead of twice as compared to conventional 6T SRAM-based memory. Also, multiple ports at the cell level include additional routing wires, increasing the memory macro's area and delay. 


A pseudo-multi-port SRAM circuit was introduced in \cite{multi}. The design is explicitly tailored for image processing tasks in display drivers. Unlike traditional SRAM designs, this circuit incorporates multiple ports in a pseudo manner. The pseudo-multi-port SRAM circuit introduces additional complexity to the memory design, requiring careful implementation and validation to ensure reliable operation. This complexity also increases the manufacturing cost, requiring additional routing for additional ports and huge area overhead. Table \ref{literature} shows the cumulative study of different multi-port memory architectures in the literature.

The major advantages of the proposed work are as follows:
\begin{itemize}
    \item A wrapper circuit is proposed to configure single-port SRAM into multi-port memory.
    \item The architecture can configure the SRAM macro as either 1-port, 2-port, 3-port or 4-port memory.
    \item The ports can be used as any combination of read and write operations.
    \item The solution enhances the memory access speed by $4\times$.
    \item The architecture provides an area-efficient solution to design multi-port memory architecture.  
\end{itemize}

\section{Proposed Multi-Port SRAM Memory Architecture}
To address the issues of area overhead and circuit complexity, we have proposed a wrapper for conventional memory architecture, which uses the conventional single-port memory architecture for configurable read/write operations from multiple ports. Fig. \ref{arch} shows the proposed multi-port architecture in which a wrapper circuit controls an SRAM macro. The wrapper circuit includes multiple input/output ports, a priority encoder, a clock generator, a finite state machine (FSM) and selection circuitry (multiplexer, decoder). Fig. \ref{arch} shows an example of 4 ports, which can be configured for both read and write operations. Each port has a port enable signal ($port\textunderscore en$), read/write enable signal ($w/rb$), address line ($addr$), and data line ($w\textunderscore data$). The ports are controlled by an external clock ($CLK$). Let us understand the functionality of each block.

\subsection{Different Blocks of Architecture}

\subsubsection{Multiple Ports}
\begin{figure}[t]
    \centering
    \includegraphics[width=\linewidth]{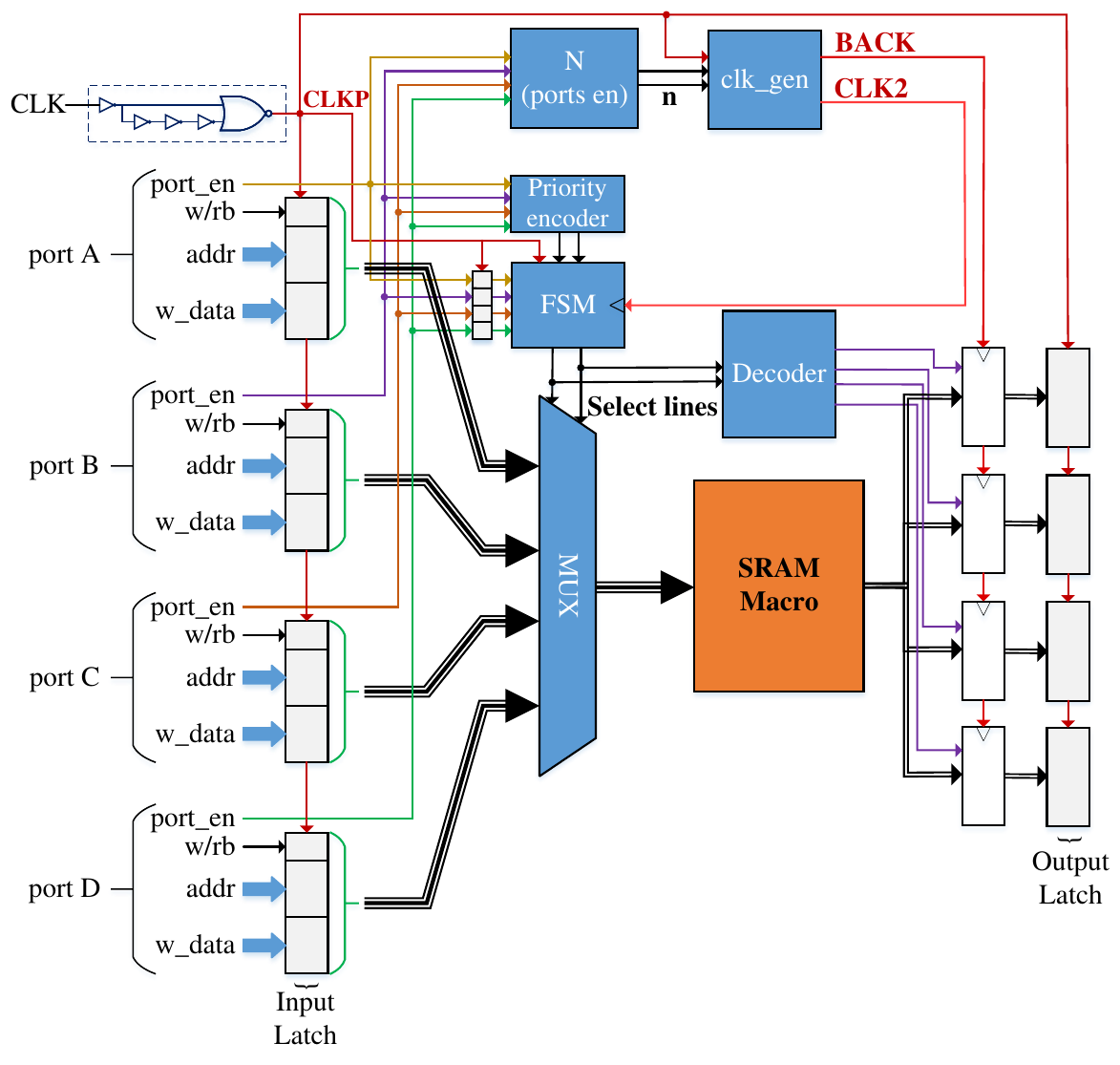}
    \caption{Proposed quad-port memory architecture. Here, SRAM macro is the conventional 6T SRAM macro}
    \label{arch}
\end{figure}

The architecture shown in Fig. \ref{arch} has four ports. Each port has one enable signal $port\textunderscore en$. By activating the $port\textunderscore en$ signal, any ports can be enabled. Based on the number of ports enabled, the architecture can be called as 1-port, 2-port, 3-port or 4-port memory. Accordingly, the $N~ports~en$ block gives the enabled port count (B1B0) to the $clk\textunderscore gen$ block, which later generates the internal clocks to switch the input/output of/from the ports.  Once the port is enabled, it must be configured for read or write operation by selecting the $w/rb$ signal. For $w/rb$ = 1, the port will perform the write operation; for $w/rb$ = 0, the port will perform the read operation. The address is given through the $addr$ signal for read or write operations. Signal $w\textunderscore data$ is used to provide the data to be written in the memory if $w/rb$ = 1.

\subsubsection{SRAM Macro}
SRAM macro consists of a 6T SRAM-based memory array. Whose memory cells are selected by the enable ports through the multiplexer. Based on the port enabled ($port\textunderscore en$) and the address ($addr$) given through the selected port, data is written or read ($w/rb$) from the respective cell.

\begin{figure}[t]
    \centering
    \includegraphics[scale=0.7]{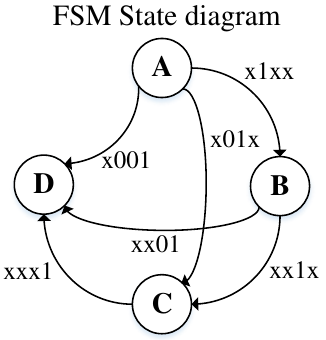}
    \caption{State diagram shows the change of states of FSM. Here, it has been considered that the priority of the ports is $A>B>C>D$. So the transition of ports starts from port A and goes till port D, later resets to port A.}
    \label{FSM}
\end{figure}

\subsubsection{Priority Encoder}
A priority encoder is used to assign priority to the enabled ports. Let us assume all four ports are enabled; priority can be given to ports, like $A>B>C>D$, based on the requirement. Based on the priority signal of the port, the FSM generates a selection signal for the MUX and transits to the next state according to the current state to transfer the data to/from the port to SRAM macro. This avoids contention issues, enhancing the responsiveness and reliability of the system.

\subsubsection{Finite State Machine}
Finite State Machine (FSM) changes the state based on port priority. Fig. \ref{FSM} shows the example of the state diagram of the FSM, where the state changes from one port to another. Transitioning between predefined states in response to inputs from multiple ports, each assigned a specific priority. When simultaneous inputs are received, the FSM evaluates the priority of each port, using priority mapping to determine the highest-priority input. The transition function then dictates the next state based on the current state, and this prioritizes input, ensuring that the most critical port connects with the SRAM macro. FSM ensures a controlled and deterministic state transition mechanism, enhancing the system's responsiveness and reliability in scenarios where multiple inputs compete for attention.

\subsubsection{Clock Generator}
\begin{figure}[t]
    \centering
    \includegraphics[width=\linewidth]{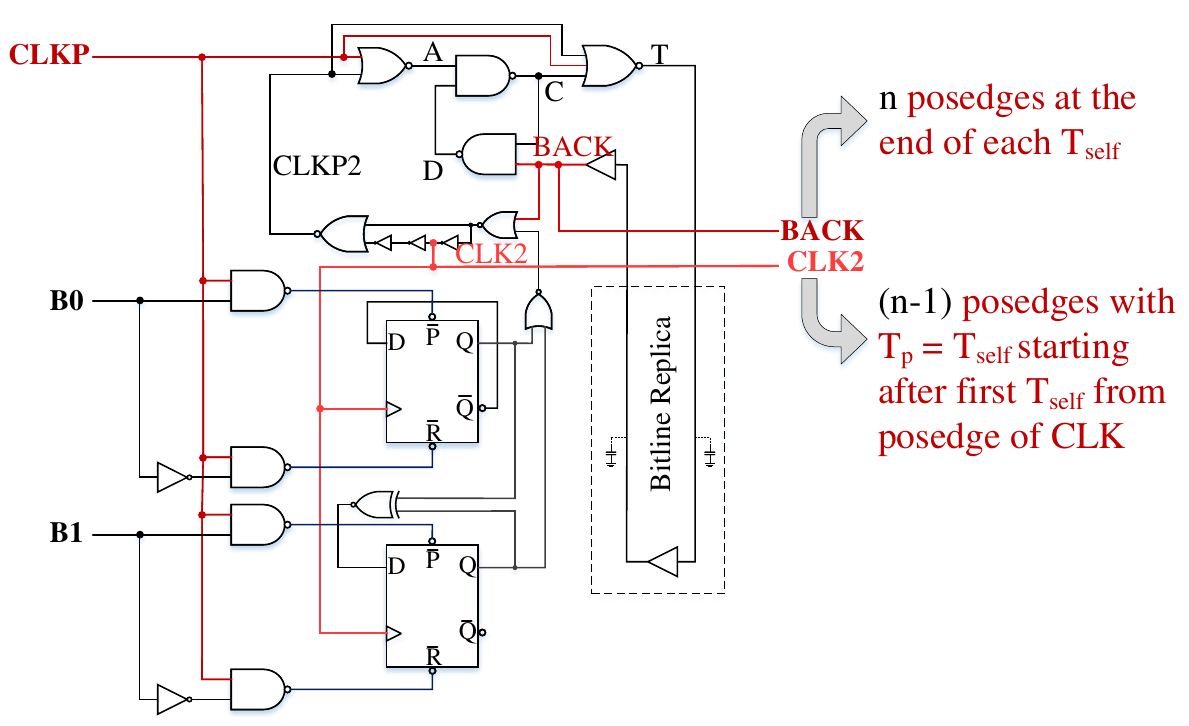}
    \caption{Block diagram of clock generator.}
    \label{clk_gen}
\end{figure}

\begin{figure}[t]
    \centering
    \includegraphics[width=\linewidth]{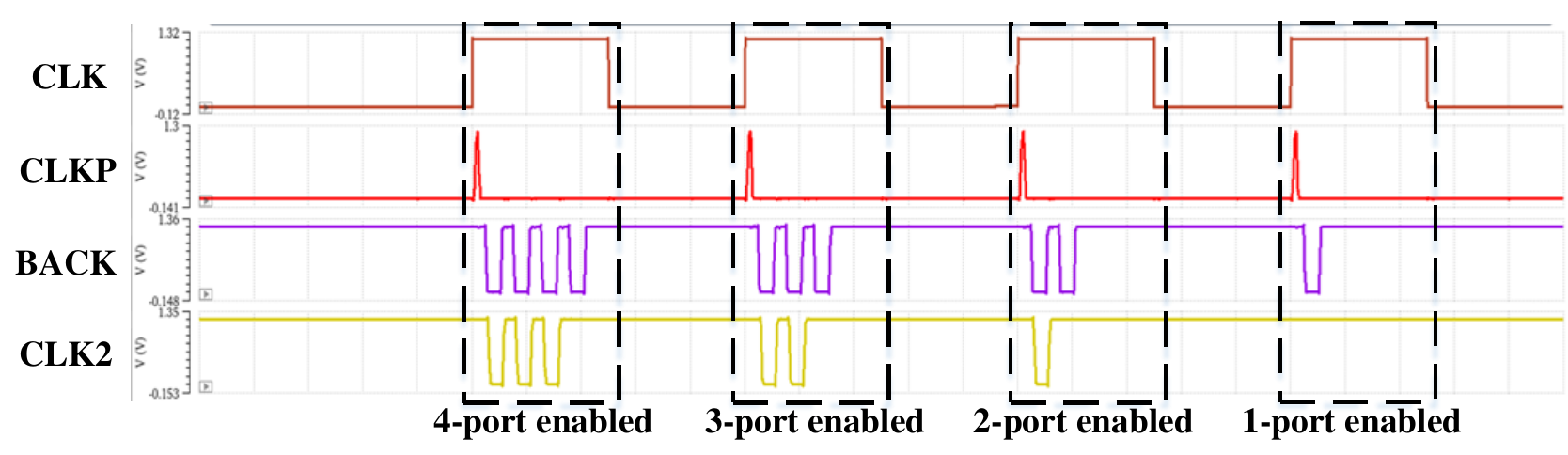}
    \caption{Simulation waveform for the clock generator. Here, simulation has been configured to 4-port, 3-port, 2-port, and 1-port memory architecture in different clocks.}
    \label{clock_gen}
\end{figure}

The clock generator is the main block of the complete wrapper, which configures the memory architecture in different port memory modes. Fig. \ref{clk_gen} shows the internal view of the clock generator block. The main function of the clock generator is to generate the $BACK$ and $CLK2$ pulses to switch the states of FSM and initialize the highest priority port after each clock $CLK$. The clock generator divides the external clock $CLK$ into multiple internal clocks based on the number of ports enabled. Fig. \ref{clock_gen} shows the simulation waveform of the clock generator. Here, the external clock $CLK$ is applied at 250MHz. The signal $CLKP$ generates the spikes at the positive edge of each clock. Based on the number of ports enabled, it divides the $CLK$ and generates $BACK$ and $CLK2$ signals. $BACK$ signal includes $N$ pulses, and $CLK2$ generates $N-1$ pulses for single $CLK$ duration. The clock generator takes the number of port-enabled counts as input B1 \& B0  (00 $\Rightarrow$ 1-port, 01 $\Rightarrow$ 2-port, 10 $\Rightarrow$ 3-port, 11 $\Rightarrow$ 4-port) and accordingly generates $BACK$ and $CLK2$ pulses. 

\subsection{Functionality of the Architecture}
The performance evolved around how data is received, and based on that, the states of the FSM changed. Let us consider all four ports are enabled, which means the architecture works as a quad-port. The inputs at ports A, B, C, and D  get latched using $CLKP$ as an input signal and select lines of the multiplexer to get initialized to port $A$. Output register corresponding to port $A$ also gets enabled to latch the read out triggered by $posedge~of~BACK$ temporarily occurs at the end of the operation cycle. In the next operation cycle, select lines change to port B, and the output register corresponding to port B gets enabled, and this process continues. As four ports are enabled, there in three ($N-1$) transitions in the state of select lines ($CLK2$) for that we get four ($N$) positive edges in $BACK$ signal. To get the ($N-1$) transitions in the select lines of the multiplexer, we require a signal $CLK2$, which is generated by a clock generator. The FSM changes its state at every $posedge~of~CLK2$ based on the input ($port\textunderscore en$) signals. The state of FSM returns back to the enabled port with the highest priority ($A>B>C>D$) at every $posedge~of~CLK$ by the asynchronous loading output of the priority encoder. At the next cycle of $CLK$, the data in the output registers got latched to the read ports.

\begin{figure}[t]
    \centering
    \includegraphics[scale=0.2]{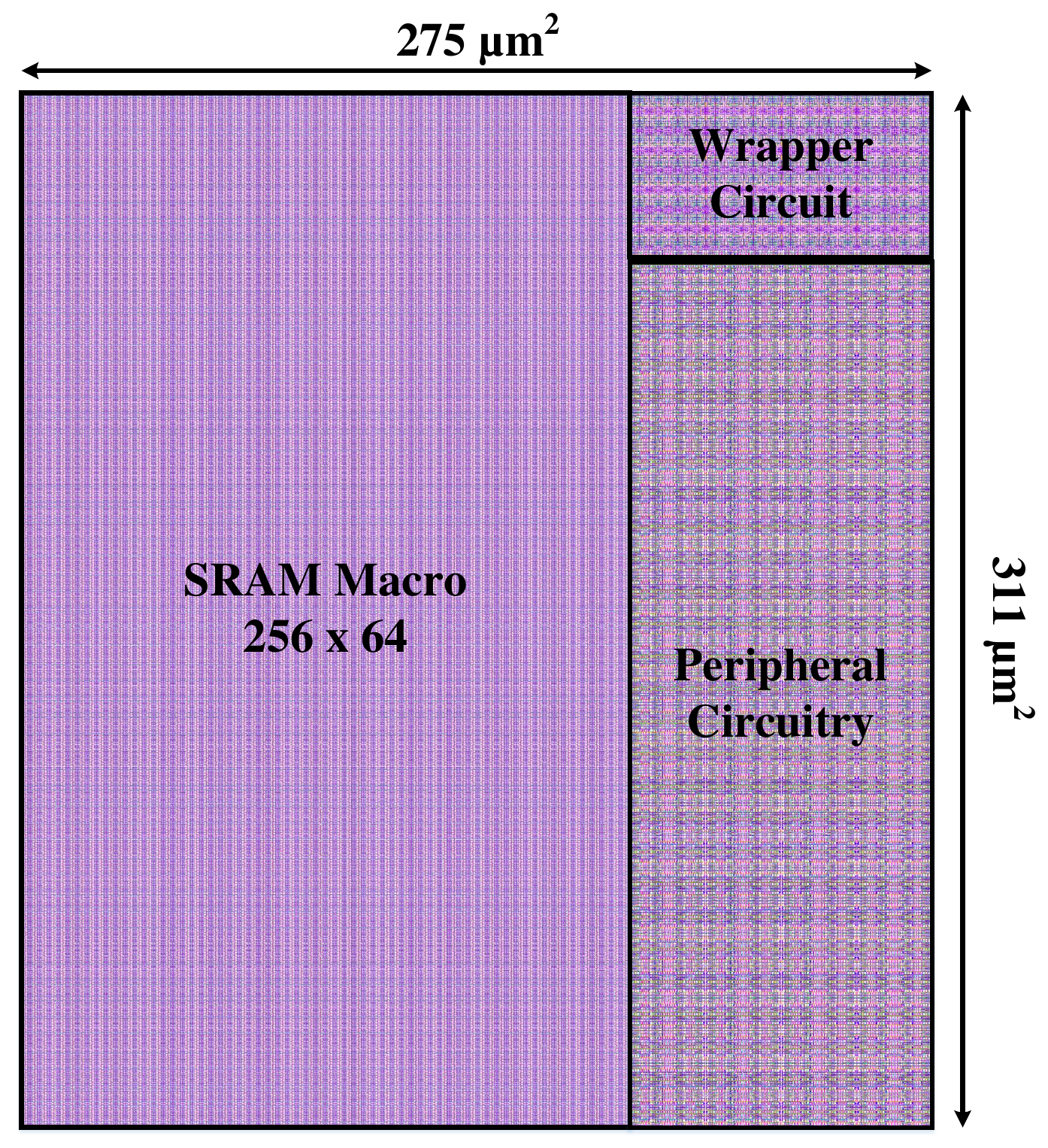}
    \caption{Layout of the overall architecture.}
    \label{layout}
\end{figure}

\begin{figure}[t]
    \centering
    \includegraphics[width=\linewidth]{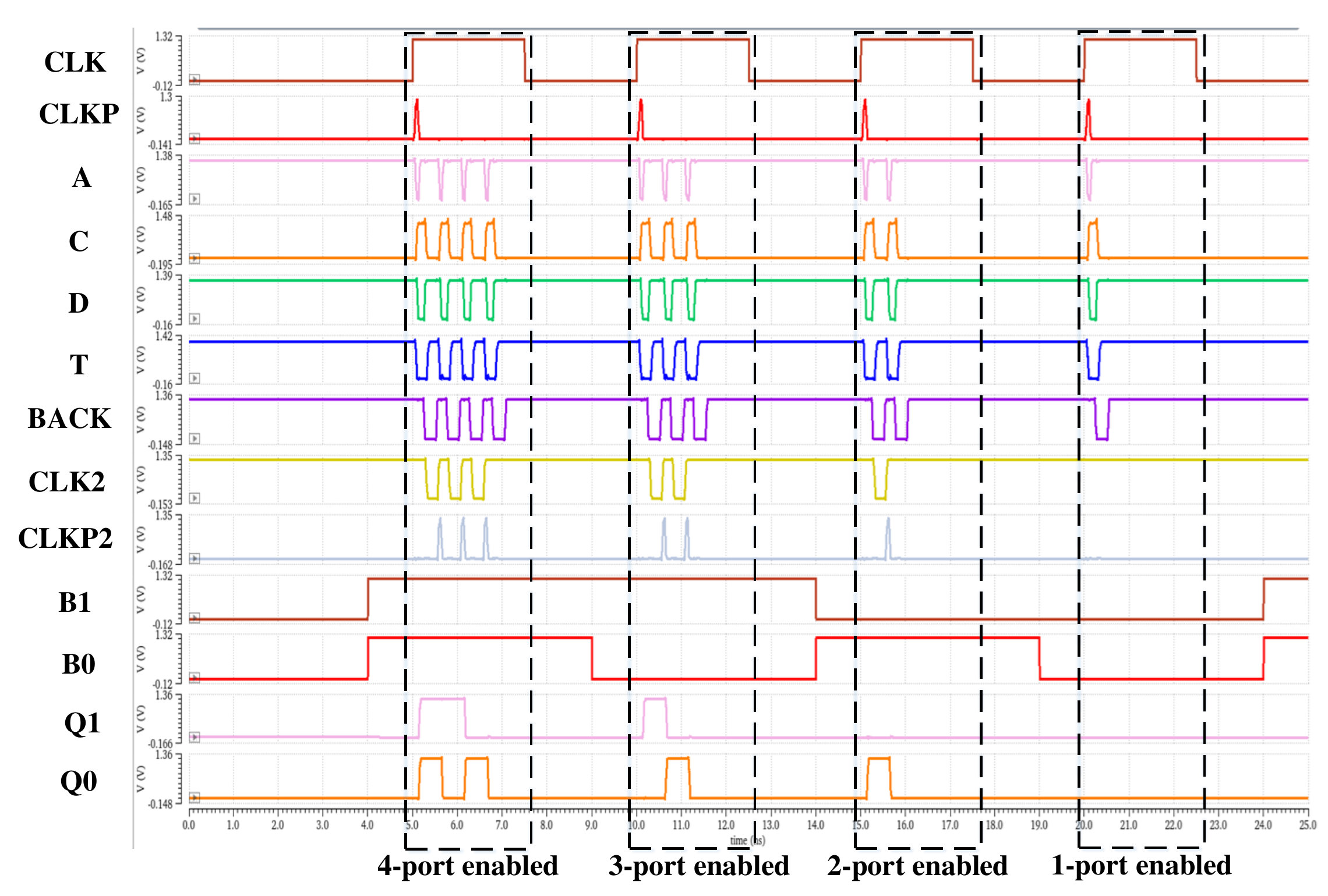}
    \caption{Simulation waveform of the overall functionality of the proposed quad-port memory architecture. Here, simulation has been configured to 4-port, 3-port, 2-port, and 1-port memory architecture in different clocks.}
    \label{P_multiport}
\end{figure}

\section{Performance Analysis}
The proposed architecture is designed with a 65nm CMOS technology node using Cadence Virtuoso and evaluated using post-layout simulations. Fig. \ref{layout} shows the layout of the overall architecture shown in Fig. \ref{arch}. Here, the wrapper circuit increases the area overhead by only 8\% for an SRAM macro of 16Kb. This is minimal compared to the area overhead due to the bitcell modification for 8T/12T/20T, etc., to add the additional ports. Fig. \ref{P_multiport} shows the simulation waveform for a quad-port-enabled system. Simulation waveforms show all the internal signals of the clock generator. Here, for each clock of $CLK$, the architecture is scheduled to work as 4-port, 3-port, 2-port, and 1-port, respectively. Here, it can be observed that in a single clock period of external clock $CLK$, the $BACK$ signal generates the $N$, and $CLK2$ generates the $N-1$ clock pulses, which is then fed to the FSM to change the state of the port. Further, it changes the states of the ports through the multiplexer, and all four ports write/read the data to/from the SRAM macro. Also, at each transition of the $CLKP$, the states were initialized to the highest priority port. Table \ref{compare} compares the proposed work with other published work. Here, the main advantage of the proposed design is that it uses an efficient 6T SRAM cell, while other work uses a higher number of transistor-based memory cells. This adds to the huge area overhead, while the area overhead of the proposed wrapper is very minimal.
The advantage of area efficiency is compared by memory density for fare comparison among different technology nodes. Table \ref{compare} also shows the improved power efficiency of the proposed architectures. Another advantage of the proposed work is that all ports can be configured for both read and write operations, while in other works, the ports were fixed for either read or write operations. The overall architecture works at the clock frequency of 250MHZ, while the memory access frequency becomes 4$\times$, i.e., 1GHz at 1.2V supply.

\begin{table*}[t]
\caption{\scshape Performance Comparison}
\centering
\begin{tabular}{|l|c|c|c|c|c|c|c|} \hline
\rowcolor{lightgray}\textbf{Parameters} & \textbf{1R1W} \cite{1R1W} & \textbf{2R2W} \cite{2R2W} & \textbf{8R1W} \cite{1W/8R} & \textbf{5R1W} \cite{5r1w} & \textbf{6R2W} \cite{6r2w}  & \textbf{6R6W} \cite{6r6w} & \textbf{Proposed} \\ \hline
\cellcolor{lightgray}Technology  & 14nm   &  6nm  &  40nm  &  90nm  &  32nm  &  7nm  & 65nm\\ \hline
\cellcolor{lightgray}Bitcell  &  8T  &  12T  &  20T  &  16T  &  24T  &  16T  & 6T    \\\hline
\cellcolor{lightgray}Bitcell Area* & 1.3$\times$ & 2$\times$ & 3.3$\times$ & 2.6$\times$ & 4$\times$ & 2.6$\times$ & 1$\times$ \\\hline
\cellcolor{lightgray}\#N Ports   & 2   & 4   & 9   & 6    & 8   & 12   & 4   \\\hline
\cellcolor{lightgray}Ports  & Fixed  & Fixed & Fixed  & Fixed &  Fixed & Fixed & Configurable\\\hline
\cellcolor{lightgray}Freq. (Memory Access)  &  2.33 GHz & -- & --  & 1.5 GHz & 250 MHz & 2.59 GHz & 1 GHz \\\hline
\cellcolor{lightgray}Supply Voltage  & 1.1 & 0.9 & --  & -- & 0.4 & 0.9 & 1.2 \\\hline
\cellcolor{lightgray}Array Size  & 72Kb & 16Kb & --  & 4Kb & 4Kb & 4Kb & 16Kb \\\hline
\cellcolor{lightgray}Area ($mm^2$)  & -- & -- & --  & 0.057 & 0.05 & 0.08 & 0.085 \\\hline
\cellcolor{lightgray}{Memory Density ($Kb/mm^2$)}  & -- & -- & --  & -- & {80} & {50} & {189} \\\hline
\cellcolor{lightgray}{Power Consumption ($mW$)} & -- & -- & --  & {5.79} & {1.2} & -- & {1.8} \\\hline
\end{tabular}\\
\vspace{0.5em}\footnotesize{*Area has been scaled with respect to 6T SRAM cell for fair comparison among different technology nodes}
\label{compare}
\end{table*}

\section{Conclusion}
This paper introduces an ultra-high-density multi-port SRAM architecture, achieved by designing a wrapper over a single-port SRAM. However, the same architecture can be scaled for any other SRAM bitcells. This wrapper circuit adds minimal area overhead compared to the significant increase resulting from modifying the memory bitcell, which typically accounts for about 60\% of the overall macro area in larger SRAM macros. The proposed circuit was implemented using 65nm CMOS technology and validated with a 1.2V supply. The wrapper retains all features of the single-port design while taking up minimal space. It allows the SRAM macro to be configured with 1-port, 2-port, 3-port, or 4-port memory, with each port flexible for use as either read or write.

\IEEEpubidadjcol


\footnotesize{
\bibliographystyle{IEEEtran}
\bibliography{Reference}} 

\end{document}